\documentclass[aps,preprint,superscriptaddress,showpacs,nofootinbib]{revtex4-1}
\usepackage{graphicx}
\usepackage{amsfonts}
\usepackage{mathrsfs}
\usepackage{epsfig}
\usepackage{slashed}
\usepackage{dcolumn}%

\usepackage{ulem}
\usepackage{color}
\usepackage{caption}
\usepackage{subcaption}
\definecolor{My_red}        {cmyk}{0.00,1.00,1.00,0.20}

\newcommand{\bmat}{\left(\begin{array}}
\newcommand{\emat}{\end{array}\right)}
\newcommand{\beq}{\begin{equation}}
\newcommand{\eeq}{\end{equation}}

\def\bwt{\begin{widetext}}
\def\ewt{\end{widetext}}
\def\be{\begin{equation}}
\def\ee{\end{equation}}
\def\bea{\begin{eqnarray}}
\def\eea{\end{eqnarray}}
\def\bean{\begin{eqnarray*}}
\def\eean{\end{eqnarray*}}
\def\bary{\begin{array}}
\def\eary{\end{array}}
\def\bit{\begin{itemize}}
\def\eit{\end{itemize}}

\def\su5u1{SU(5) \times U(1)}
\def\fsu5u1{SU(5) \times U(1)'}
\def\so10{SO(10)}
\def\sq20{SO(10) \times SO(10)}

\def\bwt{\begin{widetext}}
\def\ewt{\end{widetext}}
\def\be{\begin{equation}}
\def\ee{\end{equation}}
\def\bea{\begin{eqnarray}}
\def\eea{\end{eqnarray}}
\def\bean{\begin{eqnarray*}}
\def\eean{\end{eqnarray*}}
\def\bary{\begin{array}}
\def\eary{\end{array}}
\def\bit{\begin{itemize}}
\def\eit{\end{itemize}}

\def\su5u1{SU(5) \times U(1)}
\def\fsu5u1{SU(5) \times U(1)'}
\def\so10{SO(10)}
\def\sq20{SO(10) \times SO(10)}

\usepackage{amsmath}
\usepackage{amssymb}

\def\Zp{Z^{\prime}}

\begin{document}

\title{New signals for vector-like down-type quark in $U(1)$ of $E_6$}

\author{Kasinath Das}
\email{kasinathdas@hri.res.in}
\affiliation{Regional Centre for
Accelerator-based Particle Physics, Harish-Chandra Research Institute, HBNI, 
Chhatnag Road, Jhusi, Allahabad 211019, India}

\author{Tianjun Li}
\email{tli@itp.ac.cn}
\affiliation{CAS Key Laboratory of Theoretical Physics,
Institute of Theoretical Physics, Chinese Academy of Sciences,
Beijing 100190, P. R. China}

\affiliation{
School of Physical Sciences, University of Chinese Academy of Sciences,
Beijing 100049, P. R. China
}

\author{S. Nandi}
\email{s.nandi@okstate.edu}
\affiliation{Department of Physics and Oklahoma Center for High Energy Physics,
Oklahoma State University, Stillwater OK 74078, USA}

\author{Santosh Kumar Rai}
\email{skrai@hri.res.in}
\affiliation{Regional Centre for
Accelerator-based Particle Physics, Harish-Chandra Research Institute, HBNI, 
Chhatnag Road, Jhusi, Allahabad 211019, India}


\begin{abstract}

We consider the pair production of vector-like down-type quarks in an $E_6$ motivated model, where each of the produced 
 down-type vector-like quark decays into an ordinary Standard Model light quark and a singlet scalar. 
Both the vector-like quark and singlet scalar appear naturally in the $E_6$ model with masses at the TeV scale with a 
favorable choice of symmetry breaking pattern. We focus on the non-standard decay of the vector-like quark 
and the new scalar which decays to two photons or two gluons. We analyze the signal for the vector-like quark production 
in the $2\gamma+\geq2j$ channel and show how the scalar and vector-like quark masses can be determined at the Large
Hadron Collider.
\end{abstract}


\preprint{HRI-RECAPP-2017-010, OSU-HEP-17-04}

\maketitle

\section{Introduction}
Elementary particle physics has been at crossroads of expecting a breakthrough to understanding what lies beyond the 
Standard Model (SM) for quite some time now. The Large Hadron Collider's (LHC) discovery of the Higgs boson \cite{Chatrchyan:2012xdj,Aad:2012tfa} came 
much to the satisfaction of confirming the SM picture of electroweak interactions. Although the reported hints of several 
phenomena that would have definitely indicated of physics beyond the SM (BSM) have not survived further scrutiny by the 
LHC experiment, the diphoton excess at LHC \cite{CMS:2015dxe,TheATLAScollaboration:2015mdt,Aaboud:2016tru} brought back the attention to heavy vector-like quarks and extended 
scalar sectors amongst many other models. The SM is widely believed to be an incomplete theory due to the lack of explanation
to several outstanding issues (e.g. neutrino masses, dark matter candidate, etc.). The Grand Unified Theories (GUTs) are 
known to present novel ideas in addressing the above issues in the SM while also proposing to unify the three SM gauge 
couplings to one at a high scale. Most of the GUT models have testable consequences at the TeV scale which are in the
form of an extra gauge group such as an extra $U(1)$  and some additional new particles with heavy masses. We look at 
such an example in the $E_6$ GUT model~\cite{Gursey:1975ki} where one gets down-type vector-like (VL) fermions charged under an extra $U(1)$
gauge symmetry. In this work we focus on an interesting signal of the down-type vector-like quark (VLQ) at LHC.

Note that vector-like fermions exist in many BSM scenarios and
a lot of phenomenological studies on the down-type VLQs exist in the literatures~\cite{delAguila:1998tp}. 
The current experimental bounds on the mass of down-type VLQ are obtained under certain assumption 
of its decay modes \cite{Sirunyan:2017usq,Sirunyan:2017ezy,Aad:2015voa,Khachatryan:2015gza,Aad:2015kqa,Aad:2015mba,Aad:2011yn}. For a down-type VLQ the searches are based on the assumption that it decays to one of the SM final states 
$Zb, Wt$ and $bh$. The current experimental lower bound on the mass of the down-type vector-like quark which mixes only with the third generation quark is around 730 GeV from Run 2 of the LHC~\cite{Sirunyan:2017usq} and is around 900 GeV from 
Run 1 of the LHC~\cite{Khachatryan:2015gza}.
Similarly, the current lower bound for a vector-like quark which mixes with the light quarks is around 760 GeV 
from Run 1 of the LHC~\cite{Aad:2011yn}. 
While strong limits can be derived from these conventional search channels, the bounds get relaxed 
once new non-standard decay modes are present and start dominating over the SM channels. In this work we discuss 
a non-standard decay channel of the VLQ and about its possible signatures in a non supersymmetric version of $E_6$ model. 
A recent work discussing detailed phenomenology of vector-like quarks in $E_6$ model can be found in Ref.~\cite{Joglekar:2016yap}. In our case, we look at the VLQs and singlet scalars which are particles already present in the 
$E_6$ GUT, as discussed later. Using appropriate symmetry breaking pattern, one $U(1)$ in addition to the 
SM gauge symmetry remain unbroken at the TeV scale or even higher.  The heavy down-type quark $xd$, which is a 
color triplet and an $SU(2)$  singlet with an electric charge of $-1/3$, is pair produced dominantly from 
two gluons via strong interactions at the LHC. Also, three such $xd$ and $\overline{xd}$ quarks naturally appear in our 
model based on $E_6$ from three fermion families. A singlet scalar is also naturally present which is responsible 
in breaking the additional $U(1)$ at the TeV scale. The pattern of symmetry breaking that we shall use gives the singlet 
scalar mass which is close to the $xd$-quark mass. Our $E_6$ model will be discussed in the 
next section. The quantum numbers  of all the particles are fixed from the $E_6$ symmetry. 
The VLQ has a dominant decay mode in the non-standard form of a SM quark and the new singlet scalar which 
is the focus of this study. We discuss the phenomenology of such a scenario and on the observable signatures for the 
vector-like down-type quarks at the LHC when the singlet scalar decays to a pair of photons or a pair of gluons. 
We shall have events with dijet/diphoton resonances at the same mass and these predictions can be tested as more 
data accumulates at the upcoming 13 TeV LHC run.

This paper is organized as follows.
In Section \ref{sec:model} below, we discuss our model  and the formalism. In Section \ref{sec:phenomenology}, 
we discuss the phenomenology of our model. This gives emphasis on the prediction 
regarding the vector-like quarks through a new channel. The Section \ref{sec:summary}
contains our conclusions and discussions.

\section{ The model and formalism} \label{sec:model}

We work with an effective symmetry at the TeV scale where the SM is augmented with an extra $U(1)'$. This extra $U(1)'$ is 
a special subgroup of the $E_6$ GUT \cite{gursey, Langacker:2008yv, 
general, PLJW, Erler:2002pr, Kang:2004pp, Kang:2004ix, Kang:2009rd}. We consider the non-supersymmetric version of
$E_6$. The symmetry group $E_6$ is special in the sense that it is anomaly free, as well as has chiral fermions. Its fundamental 
representation decomposes under $SO(10)$ as

$$\bf 27 = 16  +10  +1~.$$

The representation {\bf 16} contains the $15$ SM fermions, as well as a right-handed neutrino. It 
decomposes under $SU(5)$ as 

$$\bf 16  =  10  + \bar{5} +1~.$$

And the {\bf 10} representation decomposes under $SU(5)$ as

$$\bf 10 = 5  +\bar{5}  +1 ~. $$

The {\bf 5} contains a color triplet and an $SU(2)_L$ doublet, whereas $\bf \bar{5}$ contains a color anti-triplet and 
another $SU(2)$ doublet, while the $\bf 1$ is a SM singlet. The gauge bosons are contained in 
the adjoint $\bf 78$ representation of $E_6$.

The full particle content of $\bf 27$ representation, which contains the SM fermions as well as 
extra fermions, are shown in the first two columns of Table \ref{E6Qcharge}. For three families of 
the SM fermions, we use three such $\bf 27$. The $E_6$ gauge symmetry can be broken as 
follows \cite{Group,Hewett:1988xc}
\begin{eqnarray}
E_6 \to\ SO(10) \times \ U(1)_{\psi} \to\ SU(5) \times\ U(1)_{\chi} \times\
U(1)_{\psi}~.~\,
\end{eqnarray}
The $U(1)_{\psi}$ and $U(1)_{\chi}$ charges for the $E_6$ fundamental ${\bf 27}$ representation 
are also  given in Table \ref{E6Qcharge}. 
The $U(1)'$ is a linear combination of
the $U(1)_{\chi}$ and $U(1)_{\psi}$
\begin{eqnarray}
Q^{\prime} &=& \cos\theta \ Q_{\chi} + \sin\theta \ Q_{\psi}~.~\,
\label{E6MIX}
\end{eqnarray}

The other  orthogonal linear combination  of  $U(1)_{\chi}$ and $U(1)_{\psi}$ as well as the $SU(5)$ are  
broken at a high scale. This will allow us to have a large doublet-triplet splitting scale, which prevents rapid 
proton decay if the $E_6$ Yukawa relations were enforced. 
This will require either two pairs of (${\bf 27}$, ${\bf {\overline{27}}}$)
and one pair of (${\bf 351'}$, ${\bf \overline{351'}}$) dimensional Higgs representations,
or one pair of (${\bf 27}$, ${\bf {\overline{27}}}$), ${\bf 78}$,
and one pair of (${\bf 351'}$, ${\bf \overline{351'}}$) dimensional Higgs representations
(detailed studies of $E_6$ theories with broken Yukawa relations can be 
found in~\cite{King:2005jy,Babu:2015psa}.) For our model, the unbroken symmetry at the TeV scale 
is $SU(3)_C \times SU(2)_L \times U(1)_Y \times U(1)'$.

\begin{table}[t]
\begin{center}
\begin{tabular}{|c| c| c| c| c|}
\hline $SO(10)$ & $SU(5)$ & $2 \sqrt{10} Q_{\chi}$ & $2 \sqrt{6}
Q_{\psi}$ & $4 \sqrt{15} Q$ \\
\hline
16   &   $10~ (Q_i, U_i^c, E_i^c )$ & --1 & 1  & $1$ \\
            &   ${\bar 5}~ ( D_i^c, L_i)$  & 3  & 1  & 7         \\
            &   $1 ~(N_i^c/T)$             & --5 & 1  & $-5$         \\
\hline
       10   &   $5~(XD_i,XL_i^c/H_u)$    & 2  & --2 & $-2$         \\
            &   ${\bar 5} ~(XD_i^c, XL_i/H_d)$ & --2 &--2 & $-8$ \\
\hline
       1    &   $1~ (XN_i/S)$                  &  0 & 4 & 10 \\
\hline
\end{tabular}
\end{center}
\caption{Decomposition of the $E_6$ fundamental  ${\bf 27}$
representation under $SO(10)$, $SU(5)$, and the $U(1)_{\chi}$,
$U(1)_{\psi}$ and $U(1)'$ charges.}
\label{E6Qcharge}
\end{table}
We explain our convention in some details as given in Table \ref{E6Qcharge}. Our notation is similar to 
what is used in the supersymmetric case. We have denoted the SM quark doublets, right-handed 
up-type quarks, right-handed down-type quarks, lepton doublets, right-handed charged leptons,
and right-handed neutrinos as $Q_i$, $U_i^c$, $D_i^c$, $L_i$, $E_i^c$,
and $N_i^c$, respectively. In our model, we introduce three fermionic ${\bf 27}$s,
one scalar Higgs doublet field $H_u$ from the doublet of ${\bf {5}}$ of $SU(5)$,
one scalar Higgs doublet field $H_d$ from the doublet of ${\bf {\bar 5}}$ of $SU(5)$,
one scalar SM singlet Higgs field $T$ from the singlet of ${\bf 16}$ of $SO(10)$,
and one scalar SM singlet Higgs field $S$ from the singlet of ${\bf 27}$ of $E_6$. 
Thus, similar to the fermions, all the scalars with masses at the TeV scale are coming from the 
${\bf 27}$ of $E_6$. Note that the new additional fermions from the ${\bf 27}$ with masses at the TeV 
scale are
 $N_i^c$, $XD_i$, $XL_i^c$, $XD_i^c$, $XL_i$, and $XN_i$. For details see 
 Table \ref{Particle-Spectrum}.

In our model, $S$ gives the Majorana masses to the right-handed neutrinos $N^c_i$ after $U(1)'$
gauge symmetry breaking, {\it i.e.}, the terms $S N_i^c N_i^c$ are
$U(1)'$ gauge invariant. Thus, the mixing angle  in our model is given by 
\begin{eqnarray}
\tan\theta = \sqrt{5/3}~.~\,
\end{eqnarray}

\begin{table}[t]
\begin{tabular}{|c|c|c|c|c|c|}
\hline
~$Q_i$~ & ~$(\mathbf{3}, \mathbf{2}, \mathbf{1/6}, \mathbf{1})$~ &
$U_i^c$ &  ~$(\mathbf{\overline{3}}, \mathbf{1}, \mathbf{-2/3}, \mathbf{1})$~ &
~$D_i^c$~ & ~$(\mathbf{\overline{3}}, \mathbf{1}, \mathbf{1/3}, \mathbf{7})$ ~\\
\hline
~$L_i$~ & ~$(\mathbf{1}, \mathbf{2},  \mathbf{-1/2}, \mathbf{7})$~ &
$E_i^c$ &  $(\mathbf{1}, \mathbf{1},  \mathbf{1}, \mathbf{1})$ &
~$N_i^c/T$~ &  $(\mathbf{1}, \mathbf{1},  \mathbf{0}, \mathbf{-5})$~ \\
\hline
~$XD_i$~ & ~$(\mathbf{3}, \mathbf{1}, \mathbf{-1/3}, \mathbf{-2})$~ &
~$XL^c_i,~H_u$~ & ~$(\mathbf{1}, \mathbf{2},  \mathbf{1/2}, \mathbf{-2})$~ &
~$XD_i^c$~ & ~$(\mathbf{\overline{3}}, \mathbf{1}, \mathbf{1/3}, \mathbf{-8})$ ~\\
\hline
~$XL_i,~H_d$~ & ~$(\mathbf{1}, \mathbf{2},  \mathbf{-1/2}, \mathbf{-8})$~ &
~$XN_i,~S$~ &  $(\mathbf{1}, \mathbf{1},  \mathbf{0}, \mathbf{10})$~ &
&     \\
\hline
\end{tabular}
\caption{The particles and their quantum numbers under the
  $SU(3)_C \times SU(2)_L \times U(1)_Y \times U(1)'$ gauge symmetry. Here,
  the correct $U(1)'$ charges are the $U(1)'$ charges in the above Table divided
  by $4{\sqrt{15}}$.}
\label{Particle-Spectrum}
\end{table}

The Higgs potential needed for our purpose giving rise to the extra $U(1)$ symmetry breaking is 
\begin{eqnarray}
  V = -m_S^2 |S|^2 -m_T^2 |T|^2  + \lambda_S |S|^4 + \lambda_T |T|^4
  + \lambda_{ST} |S|^2|T|^2 + (\sigma S T^2 +H.C.) ~.~\,
\end{eqnarray}
Among the parameters in the potential $V$, $\sigma$ is in general complex ($\sigma_1+i\sigma_2$) and all others are real.
Note that without the term $\sigma S T^2$, there are two global $U(1)$ symmetries for the 
complex phases of $S$ and $T$. After $S$ and $T$ obtain the Vacuum Expectation Values (VEVs), 
we have two Goldstone bosons, and one of them is eaten by the extra $U(1)$ gauge boson. Thus, to 
avoid the extra Goldstone boson, one needs the term $\sigma S T^2$ to break one global symmetry. 
This leaves us with only one $U(1)$ symmetry in the above potential, which is the extra $U(1)'$ gauge 
symmetry. Thus, after $S$ and $T$ acquire the VEVs, the $U(1)'$ gauge symmetry is broken, and 
$S$ and $T$ will be mixed via the $\lambda_{ST} |S|^2|T|^2$ and $\sigma S T^2$ terms.

The SM gauge boson masses are determined by the VEVs of the $SU(2)$ doublet scalars
and therefore $v_{EW}=\sqrt{v_d^2 + v_u^2} \simeq 246$ GeV. 
The structure for the VEVs is given as 
\begin{align}
  <H_d> &= \begin{pmatrix} v_d/\sqrt{2} \\ 0 \end{pmatrix}~,~ & <H_u> &= \begin{pmatrix} 0 \\ v_u/\sqrt{2}\end{pmatrix}~,~ \\
  <T> &= v_t/\sqrt{2}~,~ & <S> &= v_s/\sqrt{2}~ ~.
\end{align}

The mass squared matrices for the scalar sectors $(s_1,t_1)$  and $(s_2,t_2)$  are respectively given by 
 \[\mathcal{M}_{(s_1,t_1)}=
\begin{pmatrix}
2v_s^2\lambda_S-\frac{v_t^2\sigma_1}{\sqrt{2}v_s} & v_t(v_s\lambda_{ST}+\sqrt{2}\sigma_1) \\
v_t(v_s\lambda_{ST}+\sqrt{2}\sigma_1) & 2v_t^2\lambda_T
\end{pmatrix}\,\,\, \text{and} \, \, \,
 \mathcal{M}_{(s_2,t_2)}=
\begin{pmatrix}
 -\frac{v_t^2\sigma_1}{\sqrt{2}v_s} & 0\\
 0 & 0
\end{pmatrix}.
\]
$\sigma_1$ is the real part of $\sigma$ and the complex part $\sigma_2$ is assumed to be zero at tree level.
These mass matrices have been obtained from the tree-level scalar potential under the assumption that there is no mixing in the $(S,T)$ and $(H_u,H_d)$ sector.
The mass eigenstates for the CP-even sector $(s_1,t_1)$  is $s_h$ and $t_h$. The massive scalar from the CP-odd sector $(s_2,t_2)$ is represented by $a_h$. 
The relation between the gauge basis and the mass basis in for $(s_1,t_1)$ sector is given by 
\begin{equation}\label{eqn:scalar_mixing_matrix}
 \begin{pmatrix}
  s_1 \\
  t_1
 \end{pmatrix}
=
\begin{pmatrix}
 \cos\alpha & \sin\alpha \\
 -\sin\alpha & \cos\alpha
\end{pmatrix}
\begin{pmatrix}
 s_h \\
 t_h
\end{pmatrix},
\end{equation}
where the mixing angle is given by 
\begin{align*}
\sin 2\alpha=\frac{2m_{12}}{\sqrt{(m_{11}-m_{22})^2+4m_{12}^2}} 
\end{align*} 
and 
\begin{align*}
\cos 2\alpha=\frac{-(m_{11}-m_{22})}{\sqrt{(m_{11}-m_{22})^2+4m_{12}^2}},
\end{align*}
while $m_{ij}$'s are different components of the matrix $\mathcal{M}_{(s1,t1)}$. For illustration for the given parameter values
$(v_s=1.5\, \text{TeV},\lambda_S=1.78\times10^{-4}, \, \lambda_T=10^{-4},v_t=10^5, \, \lambda_{ST}=1.86\times10^{-2},\, \sigma_1=-9.8503\times10^{-1})$, we get the masses of the scalars to be 
$m_{s_h}=600$ GeV, $m_{t_h}=2507$ GeV and $m_{a_h}=2155$ GeV. The value of $\sin\alpha$ for the above parameter set is 0.82.

The Yukawa couplings in our model are 
\begin{eqnarray}
  -{\cal L} &=& y_{ij}^U Q_i U_j^c H_u + y_{ij}^D Q_i D_j^c H_d + y_{ij}^E L_i E_j^c H_d
  + y_{ij}^N L_i N_j^c H_u + y_{ij}^{XNd} XL_i^c XN_j H_d \nonumber\\&&
  + y_{ij}^{XNu} XL_i XN_j H_u 
  + y_{ij}^{TD} D_i^c XD_j {T} + y_{ij}^{TL} XL_i^c L_j {T}
\nonumber\\&&
+ y_{ij}^{SD} XD_i^c XD_j S
  + y_{ij}^{SL} XL_i^c XL_j S + {\rm H. C.}~,~\,    
  \label{lag:yukawa}
\end{eqnarray}
where $i=1,~2,~3$.
Thus,  after $S$ and $T$ obtain VEVs or after $U(1)'$ gauge symmetry breaking,
$(XD_i^c,~XD_i)$ and $(XL_i^c,~XL_i)$ will become vector-like particles
from the $y_{ij}^{SD} XD_i^c XD_j S$ and $y_{ij}^{SL} XL_i^c XL_j S$ terms,
and $(D_i^c,~XD_i)$ and $(XL_i^c,~L_i)$ will obtain vector-like masses
from the $ y_{ij}^{TD} D_i^c XD_j {T}$ and $y_{ij}^{TL} XL_i^c L_j {T}$ terms.
For simplicity, we assume $y_{ij}^{SD} \langle S \rangle >>  y_{ij}^{TD} \langle  {T}\rangle$
and $y_{ij}^{SL} \langle S \rangle >>  y_{ij}^{TL} \langle {T}\rangle$.
After we diagonalize their mass matrices, we obtain the mixings
between $XD_i^c$ and $D_i^c$, and the mixings between $XL_i$ and $L_i$.
The discussion of the Higgs potential for electroweak symmetry breaking is similar to
the Type II two Higgs doublet model, so we will not repeat it here.

We note that the $U(1)'$ gauge boson couples to all the SM fields in addition to
the new matter and scalar fields. The covariant derivatives for the $SU(2)_L$ doublet and the singlet scalars 
are respectively given by 
\begin{align}
\label{covariant}
\begin{split}
 {\mathcal{D}}_{\mu} &= (\partial_\mu -i \frac{\vec{\sigma}}{2}.\vec{W_\mu} - i g^\prime Y  B_\mu - i g_{X} Y_{X} \Zp{_\mu}),
\end{split}
\end{align}
where $Y(H_u)=\frac{1}{2}, Y(H_d)=-\frac{1}{2}$ and $Y_X(H_u)=-\frac{2}{4\sqrt{15}}, Y_X(H_d)=-\frac{8}{4\sqrt{15}}$; 
\begin{align}
\label{covariant}
\begin{split}
 {\mathcal{D}}_{\mu} &= (\partial_\mu  - i g_{X} Y_{X} \Zp{_\mu}),
\end{split}
\end{align}
where $Y_X(S)=\frac{10}{4\sqrt{15}}$ and $Y_X(T)=-\frac{5}{4\sqrt{15}}$. 
The mass square matrix for the neutral gauge boson sector in the $(W_3,B,Z^\prime)$ basis is then 
given as
\begin{align}
 \mathcal{M}=
\begin{pmatrix}
 {\Huge({\mathcal{M}_{SM}})_{2\times2}} & \begin{matrix} \mathcal{M}_{13} \\ \mathcal{M}_{23} \end{matrix} \\
\begin{matrix} \mathcal{M}_{13} & \mathcal{M}_{23}\end{matrix} & \mathcal{M} _{33}
\end{pmatrix}~,
\end{align}
where 
\begin{align}
  \mathcal{M}_{13}=\frac{gg_X}{8\sqrt{15}}(2v_u^2-8v_d^2) &,&
  \mathcal{M}_{23}=-\frac{g^\prime g_X}{8\sqrt{15}}(2v_u^2-8v_d^2)~,  \nonumber \\
{\rm and} && \mathcal{M}_{33} = \frac{g_X^2}{240}(4v_u^2+64v_d^2+25v_t^2+100v_s^2)~.
\label{eq:mzprime}
\end{align}
We can clearly see that the new gauge boson mass is dependent on the VEVs of all the scalars, 
such that one can choose one singlet VEV to be much smaller than the other and still have a very 
heavy $Z'$ that evades the existing limits. Moreover, the mixings between $W_3/B$ and $Z^\prime$
will be zero at tree level if $v_u=2v_d$.

The mass matrix for the down-type quarks and the charged leptons in the 
$(q_1,q_2,q_3,xq_1,xq_2,xq_3)$ basis is given by
\begin{align}
  \frac{1}{\sqrt{2}}
\begin{pmatrix}
y_{ij}^D v_d &  0\\
y_{ji}^{TD}v_t  & y_{ij}^{SD}v_s 
\end{pmatrix},  &&
  \frac{1}{\sqrt{2}} 
\begin{pmatrix}
y_{ij}^E v_d &  y_{ji}^{TL}v_t  \\
 0 & y_{ij}^{SL} v_s 
\end{pmatrix},
\end{align}
where $i,j=1,2,3$. The $q_i$s and $xq_i$s represent the down-type quarks for the left matrix and charged 
leptons for the right matrix. These mass matrices would be diagonalized by a bi-unitary 
transformation which would 
lead to a mixing between the vector-like fermions and the SM fermions. However, one should note that 
the mixings between the left-handed fermions and the right-handed fermions will be dictated by  different
set of mixing angles. In our analysis we will allow mixings between the $d$ quark and the 1st generation vector-like quark($xd_1^0$) only and the mass matrix 
in the gauge basis $(d^0,xd_1^0)$  is given by 
\[
\frac{1}{\sqrt{2}}
 \begin{pmatrix}
  y_{11}^Dv_d & 0\\
  y_{11}^{TD}v_t & y_{11}^{SD}v_s
 \end{pmatrix}\equiv
 \begin{pmatrix}
  m_1 & 0\\
  m_2 & m_3
 \end{pmatrix}.
\]
The mixing matrices which transform the gauge eigenstates $(d^0,xd_1^0)$ to mass eigenstates$(d,xd_1)$ are given by
\begin{equation}\label{eqn:quark_mixing_matrix}
 S_i = 
 \begin{pmatrix}
  \cos\theta_i & -\sin\theta_i \\
  \sin\theta_i & \cos\theta_i
  \end{pmatrix},
  \text{where}\,\,\, i = L,R,
\end{equation}

with the following left and right handed mixing angles
\begin{align*}
\sin 2\theta_L=\frac{2m_1m_2}{\sqrt{(m_1^2-m_2^2-m_3^2)^2+4m_1^2m_2^2}}, \,\,\,\,\,\,\,\,\, &
\cos 2\theta_L=\frac{-(m_1^2-m_2^2-m_3^2)}{\sqrt{(m_1^2-m_2^2-m_3^2)^2+4m_1^2m_2^2}},
\end{align*}

\begin{align*}
\sin 2\theta_R=\frac{2m_2m_3}{\sqrt{(m_1^2+m_2^2-m_3^2)^2+4m_2^2m_3^2}}, \,\,\,\,\,\,\,\,\,\, &
\cos 2\theta_R=\frac{-(m_1^2+m_2^2-m_3^2)}{\sqrt{(m_1^2+m_2^2-m_3^2)^2+4m_2^2m_3^2}}.
\end{align*}

We should also point out a few useful assumptions that we think are relevant for the analysis:

\begin{enumerate}
  \item We have neglected any mixing between the electroweak doublet scalars and singlet scalars. 
  \item We also ensure that the new $U(1)'$ gauge boson does not have a significant mixing with the 
  SM gauge boson $Z$ ($\mathcal{M}_{13}, \mathcal{M}_{23} << \mathcal{M}_{33}$). 
  \item For simplicity, we will take all types of  Yukawa couplings $y^A_{ij}$   
to be zero for $i\neq j$, where $A\equiv TD, TL, SD, SL$ (see eq.\ref{lag:yukawa}).
  \item The mixing angles between the left-handed SM fermions and the vector-like fermions are taken to be very very small, {\it i.e.},
  we assume the  left-mixing angle $\theta_L\sim0$  to avoid the flavour physics constraints~\cite{Alok:2014yua}. For the choice of the set of parameter values 
  \{$\frac{y_{11}^Dv_d}{\sqrt{2}}\sim m_d$, $\frac{y_{11}^{SD}v_s}{\sqrt{2}}\sim640$ GeV, $v_t\sim10^4$ GeV, $y_{11}^{TD}\sim10^{-5}$\} we get small
  values of mixing angles, {\it i.e.}, $\sin\theta_L\sim10^{-10}$ and $\sin\theta_R\sim10^{-4}$.
  
\end{enumerate}

\section{Signals for vector-like quarks}\label{sec:phenomenology}
\begin{figure}[h!]
\includegraphics[width=15cm]{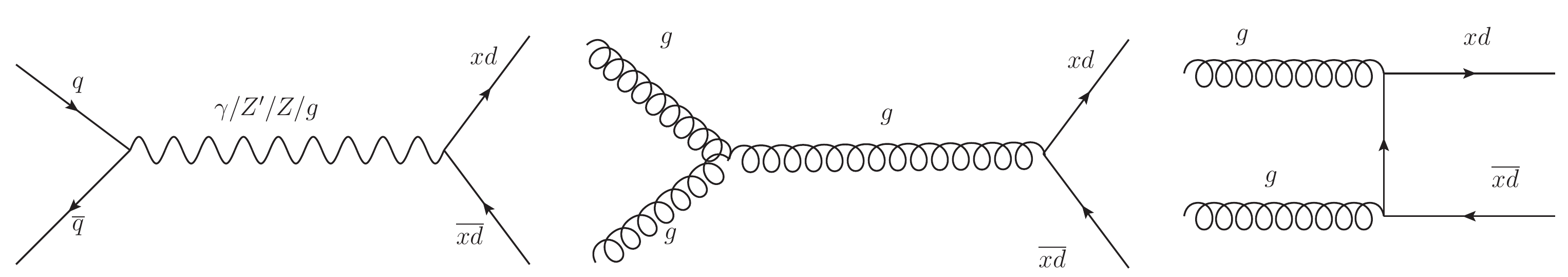}
\caption{The leading-order Feynman diagrams for the subprocess contributing to the pair production of the VLQ at the LHC.}
\label{fig:feynman1}
\end{figure} 
The new VLQ will be dominantly produced via strong interaction, with subleading contributions coming from the $s$-channel 
exchange of the $\gamma, Z$ and $Z^\prime$. In situations where the VLQ mass is less than $M_{Z'}/2$, then the 
$Z^\prime$ mediated process can give a resonant contribution. However these contributions are found to be not very significant.
We list the various production mechanisms of the VLQ in Fig.\ref{fig:feynman1}. 
Note that one can in principle also produce the VLQs singly but they would be heavily suppressed as the production 
strength would depend on the mixing between the VLQs and SM quarks.

\begin{figure}[h!]
\includegraphics[width=3.5in,height=3.5in]{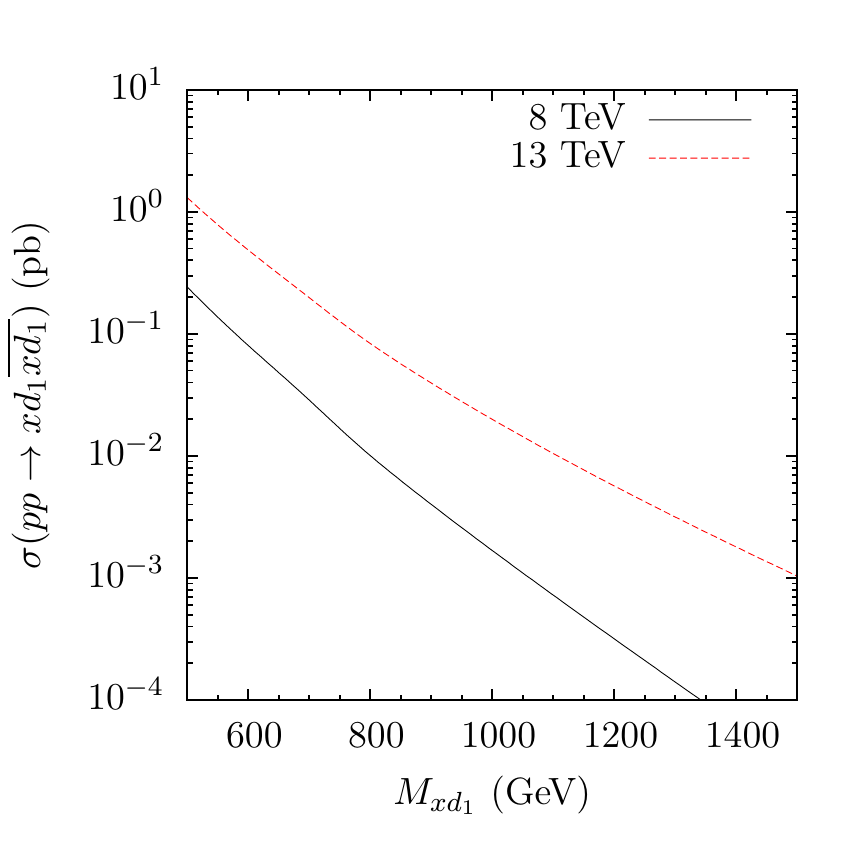}
\caption{The pair production cross section of $xd_1$ at the LHC 
with $\sqrt{s}=13$ TeV as a function of $M_{xd_1}$.}
 \label{fig:xd_production}
\end{figure}

In Fig.~\ref{fig:xd_production} we show the pair production cross section of the VLQ $xd_1$ as a function of its mass at both run-1
and current run of the LHC with $\sqrt{s}=13$ TeV. With a few 100 femtobarns of cross section, it would be highly unlikely for 
the LHC to 
miss the signal for VLQs when they decay directly to SM particles. These already put strong limits on the mass of the VLQs. 
However,  a new decay mode for the VLQ can definitely alter the search strategies for these exotics even when the rates are 
significantly high.

With the details of the model discussed in the previous section, it is now possible to write down the interaction vertices 
for the VLQ and new scalars with the SM particles that we use in our calculation and analysis. We list the relevant interactions 
in Table \ref{table:couplings}. 

\setlength{\tabcolsep}{10pt}
\begin{table}
 \begin{tabular}{|c  |c|c|c|}
 \hline
  & \boldmath{$K$} & \boldmath{$c_V$} & \boldmath{$c_A$} \\
  \hline
    $\overline{d} \, d \, Z_\mu$ & $\frac{e}{12\sin{2\theta_w}}$ & $4\cos{2\theta_w}+3\cos{2\theta_L}-1$ & $3(1+\cos{2\theta_L})$\\
\hline
  $\overline{xd_1} \, xd_1 \, Z_\mu$ & $\frac{e}{12\sin{2\theta_w}}$ & $4\cos{2\theta_w}-3\cos{2\theta_L}-1$ & $3(1-\cos{2\theta_L})$\\
  \hline
  $\overline d \, xd_1 \, Z_\mu$ & $\frac{e}{4}\frac{\sin{2\theta_L}}{\sin{2\theta_w}}$ &1 &1\\
  \hline
  $\overline{u} \, d\, W^+$ & $\frac{-e}{2\sqrt{2}}\frac{\cos{\theta_L}}{\sin{\theta_w}}V_{ud}$ & 1&1\\
  \hline
    $\overline{u}\, xd_1 \, W^+$ & $\frac{-e}{2\sqrt{2}}\frac{\sin{\theta_L}}{\sin{\theta_w}}V_{ud}$ & 1&1 \\
    \hline
    $\overline{d} \, d \, Z'_\mu$ & $\frac{g_X}{16\sqrt{15}}$ & $15\cos{2\theta_R}-3\cos{2\theta_L}$ & $-(15\cos{2\theta_R } +3\cos{2\theta_L}-2)$ \\
    \hline
    $\overline{xd_1} \, xd_1 \, Z'_\mu$ & $-\frac{g_X}{16\sqrt{15}}$ & $15\cos{2\theta_R}-3\cos{2\theta_L}$ & $-(15\cos{2\theta_R } +3\cos{2\theta_L}+2)$ \\
    \hline
    $\overline{d} \, xd_1 \, Z'_\mu$ & $\frac{3 g_X}{16\sqrt{15}}$ & $5\sin{2\theta_R}-\sin{2\theta_L}$ & $-(5\sin{2\theta_R } +\sin{2\theta_L})$ \\
    \hline\hline
  
  & \boldmath{$K$} & \boldmath{$c_S$} & \boldmath{$c_P$}\\
  \hline\hline
  $\overline{d} \,d \,s_h$ & $-\frac{\sin{\theta_L}}{\sqrt{2}}$& $y_{11}^{SD}\cos{\alpha}\sin\theta_R+y_{11}^{TD}\sin\alpha\cos\theta_R$ &0\\
  \hline
  $\overline{xd_1} \, xd_1 \, s_h$& $\frac{-\cos\theta_L}{\sqrt{2}}$& $y_{11}^{SD}\cos{\alpha}\cos\theta_R-y_{11}^{TD}\sin\alpha\sin\theta_R$&0\\
  \hline
  & & $y_{11}^{SD}\cos\alpha\sin(\theta_L+\theta_R)$ &$y_{11}^{SD}\cos\alpha\sin(\theta_L-\theta_R)$\\
 \raisebox{1.7ex}{$\overline{d} \, xd_1 \, s_h$} & \raisebox{1.7ex}{$\frac{1}{2\sqrt{2}}$} & $+\,y_{11}^{TD}\sin\alpha\cos(\theta_L+\theta_R)$ & $-\,y_{11}^{TD}\sin\alpha\cos(\theta_L-\theta_R)$\\
 \hline
 $\overline{d} \, d \, t_h$&$\frac{-\sin\theta_L}{\sqrt{2}}$ &$y_{11}^{SD}\sin\alpha\sin\theta_R-y_{11}^{TD}\cos\alpha\cos\theta_R$&0 \\
 \hline
   & & $y_{11}^{SD}\sin\alpha\sin(\theta_L+\theta_R)$ &$y_{11}^{SD}\sin\alpha\sin(\theta_L-\theta_R)$\\
 \raisebox{1.9ex}{$\overline{d} \, xd_1 \, t_h$} & \raisebox{1.9ex}{$\frac{1}{2\sqrt{2}}$} & $-\,y_{11}^{TD}\cos\alpha\cos(\theta_L+\theta_R)$ & $+\,y_{11}^{TD}\cos\alpha\cos(\theta_L-\theta_R)$\\
 \hline
 $\overline{xd_1} \, xd_1 \, t_h$& $\frac{-\cos\theta_L}{\sqrt{2}}$& $y_{11}^{SD}\sin\alpha\cos\theta_R\,+\,y_{11}^{TD}\cos\alpha\sin\theta_R$&0\\
 \hline
 \end{tabular}
\caption{The couplings of VLQ $xd_1$ and $d$-quark with SM gauge bosons and, with the scalars $s_h$ and $t_h$. 
Coupling with gauge bosons are of the form  $K\gamma^{\mu}(c_V-c_A \gamma^{5})$ and with scalars are of the 
form $K(c_S-c_P\gamma^{5})$. Here $\alpha$ is the  scalar sector mixing angle in Eq.~(\ref{eqn:scalar_mixing_matrix}), 
$\theta_L$ and $\theta_R$ are left and right
mixing angles in Eq.~(\ref{eqn:quark_mixing_matrix}).}
\label{table:couplings}
\end{table}
\setlength{\tabcolsep}{6pt}

The possible decay modes for a down-type VLQ in our model are to the SM particles 
given by $xd_1 \rightarrow h \,d, \,d \,Z, \, \text{and} \, u \,W^{-}$ while the non-standard decay modes 
would be $ xd_1 \rightarrow s_h\,d, \,t_h \,d, \,a_h \,d, \text{and} \,d \,Z^\prime$. 
Here $h$ is the SM Higgs boson, $u$ and $d$ are SM first generation quarks, $Z$ and $W^-$ are SM 
gauge bosons. $s_h$ and $t_h$ are CP-even scalars from $(S,T)$ sector while 
$a_h$ is CP-odd scalar from $(S,T)$ sector.
For simplicity we focus on the case where out of all the SM down-type quarks, $xd_1$ interacts only with the $d$ quark through the Yukawa interaction with $S$ and $T$. For a very small mixing between  $xd_1$ and $d$ quark and for a $Z'$ heavier than $xd_1$, 
the dominant decay modes of $xd_1$ become $s_h \, d, \,\, t_h\,d$ and $a_h\,d$. 
The other decay modes are suppressed because the interaction strength for these 
decays are proportional to the very small mixing angles $\sin\theta_L$ and 
$\sin\theta_R$ (Table \ref{table:couplings}). 
As discussed in the previous section and to be safe from flavor constraints, one 
can impose small mixing angles, for example, $\sin\theta_L\sim10^{-10}$ and 
$\sin\theta_R\sim10^{-4}$ as mentioned for a set of parameter choices of the model.
This will insure that the vector-like fermions do not   
decay to the SM gauge bosons and light SM fermions~\cite{Grossmann:2010wm}. The decay to the SM Higgs and light down-type quark is again
very suppressed, due to the coupling strength being proportional to $\sin\theta_R$ and mass of the down-type SM quark.
The mixing in the Higgs sector has been neglected as a convenient choice to keep the 
number of free parameters to tune to be small.
  
The $Z-Z^\prime$ mixing which is anyhow strongly constrained by electroweak data in any 
$U(1)$ extension beyond the SM is also negligible, 
thus avoiding the decay of VLQ to $d\,Z$ final state through this mixing.
All other possible scalars other than $h$ from the doublet sector $(H_u, \, H_d)$ are 
heavier than the VLQ, and thus ensure absence of the decay of VLQ to them.
So if both $a_h$ and $t_h$ are also heavier than $xd_1$, the VLQ $xd_1$ 
decays to the lone $s_h \,d$ final state. This decay is not suppressed due to a direct Yukawa coupling of the SM quark and VLQ with $T$ 
as well as the mixing between the $xd_1$ and $d$ in the right-handed sector. Thus, the decay is made possible through not only 
the mixing between the CP-even components of the scalars $S$ and $T$ but also depends on $\sin\theta_R$.  
It turns out that even with a choice of the Yukawa strength of 
$10^{-5}$ or lower (where $\sin\theta_L\sim10^{-10}$ and $\sin\theta_R\sim10^{-4}$), 
this decay is still the dominant channel. Thus, with 
the minimal assumptions that mixing of the new states with the SM sector being small and negligible 
allows a very specific decay channel for the VLQ in the model. 

 Once the VLQ is produced at the LHC, it will almost always decay into the non-standard channel to 
 give a light quark jet and the scalar $s_h$. The $s_h$ then decays 
promptly to either SM particles or any lighter states of the new particles in the spectrum.
The decay modes for the scalar can be summarized as $s_h \rightarrow 
\overline{\ell_i} \,\ell_j, \overline{x\ell_i} \,x\ell_j, \,\overline{x\ell_i} \, \ell_j, \, \overline{\ell_i} \,x\ell_j, \, \gamma \, \gamma, \, g \,g, 
\overline{d} \, d$. 
Here $\ell_i$s are SM charged leptons and $x\ell_i$s are vector-like leptons. We avoid the decay 
of $s_h$ to a pair of VLLs  by setting their mass such that $M_{x\ell_i}>m_{s_h}/2$. Here $M_{x\ell_i}$ is 
mass for the $i^{th}$ generation vector-like lepton and $m_{s_h}$ is the mass of the scalar $s_h$. 
The Yukawa coupling $y_{ij}^{TL}$  has been chosen zero to avoid mixing between the VLL and the SM 
lepton sector, and thus avoiding the decay of $s_h$ to the final states $\overline{x\ell_i} \, \ell_j$, 
$\bar{\ell_i} \, x\ell_j$ and $\overline{\ell_i} \, \ell_j$. Additionally the decay to $\overline{d}d$ is controlled 
by the mixing angle $\sin\theta_L$ and is therefore too suppressed. 
With $xd_1$ heavier than $s_h$ the decay mode 
$s_h \rightarrow\overline{xd_1}\,d$ is not allowed. Hence the only allowed final states for $s_h$ decay are 
$\gamma\gamma$ and $gg$.

Decay to $\gamma\gamma$ and $gg$ will occur through the 
effective one-loop induced coupling.
All of the three generations of down-type VLQs and charged VLLs will affect the branching ratios of $s_h$ to $\gamma\gamma$ and to $gg$. The coupling of $s_h$ to gluons and photons follows the standard notations that are being used in the literature and 
for clarification we are giving only the 
coupling with gluons  by the effective 
Lagrangian
\begin{align} \label{eq:sgg}
\mathcal{L}_{s_hGG} = - \lambda_{sgg} s_h ~G_{\mu\nu} G^{\mu\nu}~,
\end{align}
with the effective coupling $\lambda_{sgg} = \alpha_s  F_{1/2} (\tau_{xd}) /( 16 \pi v_s) $ where
\begin{align}
  F_{1/2}(\tau_{xd}) = 2(\tau_{xd}+(\tau_{xd}-1)f(\tau_{xd}))\tau_{xd}^{-2}
\end{align} 
represents the loop function and 
$f(\tau_{xd})=(sin^{-1}\sqrt{\tau_{xd}})^2$ with $\tau_{xd}=m_{s_h}^2/4M_{xd}^2 \,\, < 1$. 
Here, we have shown the contribution to the coupling from only one vector-like quark.

 \begin{figure}[h!]
\includegraphics[width=3.0in,height=2.8in]{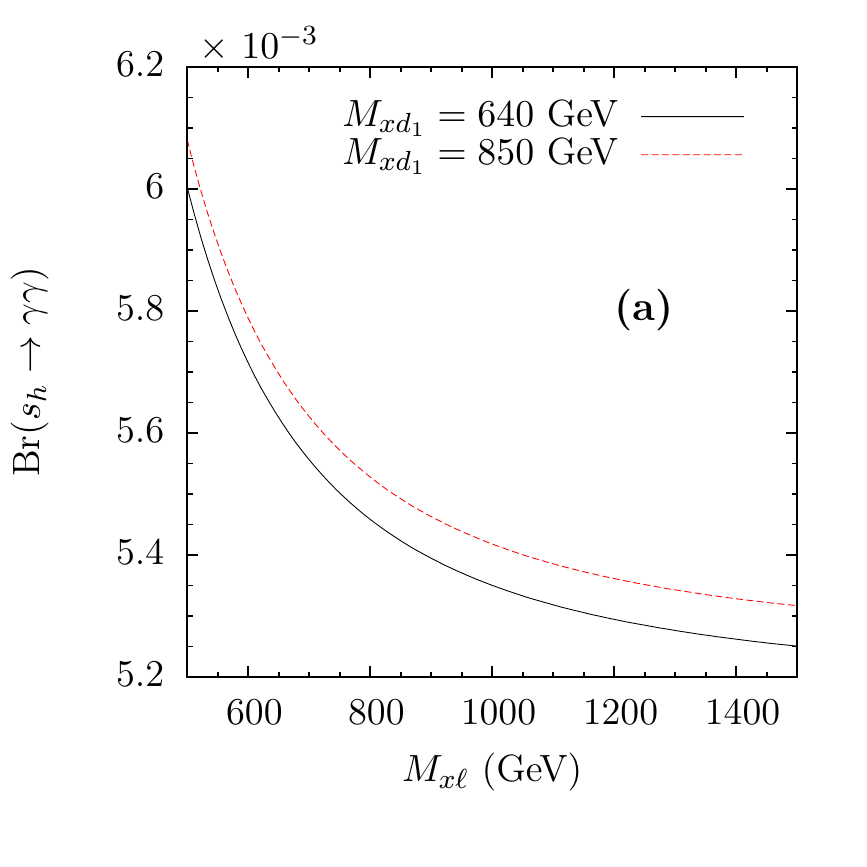}
 \includegraphics[width=3.0in,height=2.9 in]{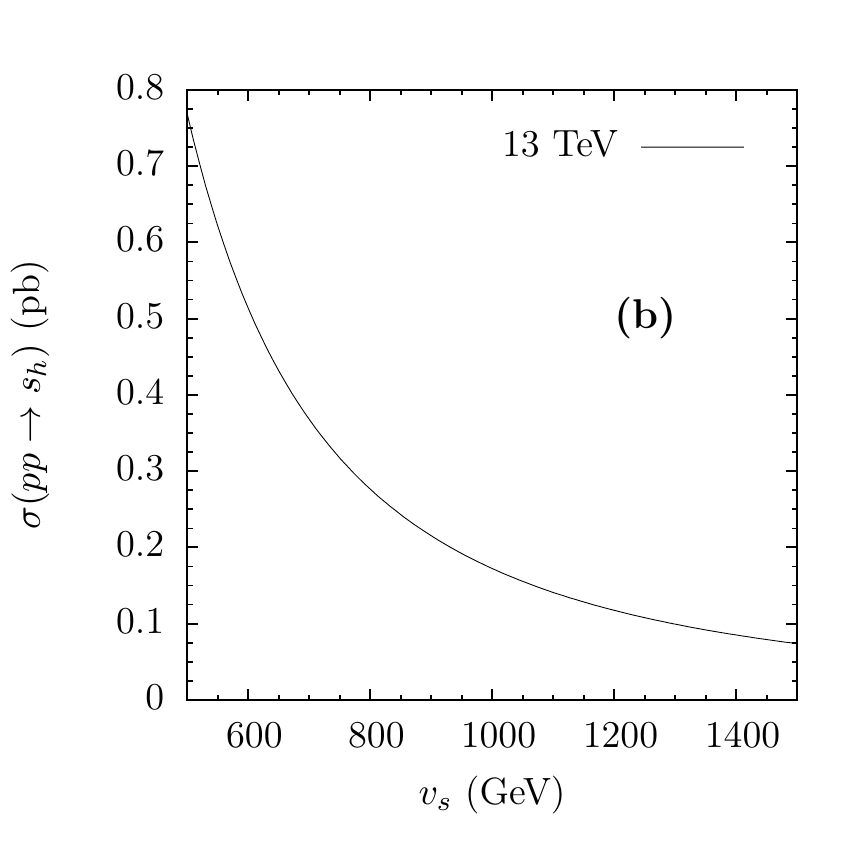}
\caption{ (a) Illustrating the diphoton branching ratio for $s_h$ decay as a function of the vector-like lepton mass for two different values of the lightest VLQ
mass ($M_{xd_1}$). 
(b) The on-shell $s_h$ production cross section at LHC with $\sqrt{s}=13$ TeV through gluon-fusion as a function the singlet vev $v_s$  with $m_{s_h}=600$ GeV. 
For the above plots, we fix $M_{xd_2}\,=\,M_{xd_3}\,=$ 1.5 TeV, $y_{ii}^{SL}\,=\,1$.  For (a) $M_{x\ell_1}\,=\,M_{x\ell_2}\,=\,M_{x\ell_3}\,=\,M_{x\ell}$, while
for (b) $y_{11}^{SD}\,=\,1$, $M_{xd_1}=y_{11}^{SD}v_s/\sqrt{2}$, and  $y_{22}^{SD}\,=\,\sqrt{2}\,M_{xd_2}/v_s$.}
 \label{fig:braa_ppsh}
\end{figure}

 We plot the branching ratio for the scalar $s_h$ decaying into a pair of photons in Fig.~\ref{fig:braa_ppsh} (a)
as a function of the VLL masses for two values of the lightest VLQ mass($M_{xd_1}$), while the other two VLQs have masses at 1.5 TeV.
As the mass of light vector-like leptons are not severely constrained by experiments, we shall consider the results with all the three VLLs contributing to the diphoton decay.  
Here the mass values  of all the three vector-like leptons have been taken to be 
same ($M_{x\ell}$) while the Yukawa couplings of $s_h$ to VLLs have been taken to 
be unity ($y_{ii}^{SL}=1$).

Note that the branching of the $s_h \to \gamma \gamma$ is very similar to 
the order at which the SM Higgs decay happens but slightly higher. This is because of the 
contributions of the VLLs which do not contribute to the $s_h \to gg$ mode. However, the decay to
$\gamma \gamma$ mode is still between 0.5\% -- 0.6\% at best while the remaining 
decay probability is made up by the $gg$ channel.

Fig.~\ref{fig:braa_ppsh} (b) shows the production cross section of a 600 GeV $s_h$ at $\sqrt{s}=13$ TeV 
center of mass energy as a function of the singlet vev $v_s$. The $s_h$ is being produced 
by the loop induced effective coupling in Eq.~\ref{eq:sgg}. For the cross section estimates we have 
chosen the Yukawa coupling of $s_h$ to the 1st generation VLQ $xd_1$ to be 1. The mass of $xd_1$ 
depends on the value $v_s$ following the relation $M_{xd_1}=y_{11}^{SD}v_s/\sqrt{2}$. The masses for 
the $xd_2$ and $xd_3$ have been taken to 1.5 TeV. The Yukawa couplings ($y_{22}^{SD}$ 
and $y_{33}^{SD}$) have different values for different values of $v_s$. Note that the $s_h$ production
crucially depends on the Yukawa coupling which can be tuned to control its production rate. In fact, this 
possibility was earlier used by us \cite{Das:2015enc,Das:2016xuc} to arrive at the possible 
explanation of the now hitherto disproved diphoton excess for a 750 GeV resonance \cite{CMS:2015dxe,TheATLAScollaboration:2015mdt,Aaboud:2016tru}.

 As pointed out earlier, the VEVs for $S$ and $T$ which are given by $v_s$ and $v_t$ respectively 
play a significant role in giving mass to $Z^\prime$. We 
choose the mass of $Z^\prime$ to be 1.5 TeV which is still allowed by current LHC data, primarily 
due to the SM fermions carrying very suppressed quantum charges of the new $U(1)^\prime$ as shown in 
Table \ref{Particle-Spectrum}. As can be seen from the mass square matrix of the neutral gauge 
boson sector (Eq.~(\ref{eq:mzprime})), by choosing large values of $v_s$ or $v_t$, it is possible 
to avoid mixing between the SM and new gauge boson sector. But the vector-like lepton masses are 
given by the relation, $M_{x\ell_i}=y_{ii}^{SL}v_s/\sqrt{2}$. And $y_{ii}^{SL}$ is the Yukawa coupling 
of $s_h$ to the VLLs which enters in the $s_h\gamma\gamma$ effective coupling. For a sub-TeV vector-like 
lepton and with significant value of Yukawa coupling it will not be possible to choose 
a very large value of $v_s$. We therefore choose a higher value for  $v_t$ at  10 TeV 
which effectively suppresses any significant mixing in the neutral gauge boson sector.

Note that after the decay of the VLQs there will be two $s_h$ and two jets in the final state. As the 
$s_h$  decays to two gluons or to two photons only, with almost 99\% to the gluonic jets,  
the resultant final states are either $2\gamma+4j$, $4\gamma+2j$ or $6j$. The cross section for 
the $4\gamma\,+\,2j$ final state is quite small while the QCD background for
the $6j$ final state is significantly large compared to the $2\gamma\,+\,4j$ final state. Thus, these two channels 
would require large statistics to leave any imprint of their signal at the LHC. So in all likelihood the remaining channel 
of $2\gamma+4j$  seems the most promising channel which we shall focus on for our analysis.

For the $2\gamma+4j$ final state one of the $xd_1$ will eventually have a full hadronic decay to 3 jets. The bound on the branching ratio for the decay of a 
color triplet vector-like quark to three jets for different masses of vector-like quark has been obtained in \cite{Dobrescu:2016pda} using the existing searches for resonances 
in multijet final states by the CDF Collaboration \cite{Aaltonen:2011sg} at the Tevatron, and by the CMS and ATLAS Collaborations at the LHC using data from the 
7 TeV \cite{ATLAS_CMS_7TeV}, 8 TeV \cite{ATLAS_CMS_8TeV} and 13 TeV \cite{ATLAS_13TeV} runs.  Ref.~\cite{Dobrescu:2016pda} shows that the physical region 
where BR(VLQ $\rightarrow 3j)\leq 1$ for different mass values of vector-like quark remains unconstrained except for a tiny region around 500 GeV.

\begin{table}[h!]
\centering
  \begin{tabular}{|l|m{5cm}|m{3cm}|m{2.4cm}|r|}
   \hline
       & Model Parameters & Particle Mass & (Br($s_h\rightarrow \gamma \gamma$), \,Br($s_h\rightarrow gg$)) & $\sigma(pp\rightarrow xd_1 \overline{xd_1})$\\
      \hline
 BP1 & $\lambda_T=0.31$,  $\lambda_S= 0.088$,  $\lambda_{ST}=0.75$,  $v_s = 950$ GeV, $v_t =10^4$ GeV, $\sigma_1= -83.96$ GeV, $\sin\alpha=0.71$ & $M_{xd_1}=640$ GeV,  $M_{x\ell_i}=500$ GeV, $m_{s_h}=600$ GeV,  $m_{a_h}=2.5$ TeV, $m_{t_h}=3.5$ TeV & (0.006,  0.994) & 339 fb \\
    \hline
  BP2 & $\lambda_T=0.01$,  $\lambda_S= 0.269$,  $\lambda_{ST}=0.216$,  $v_s=2.5$ TeV, $v_t =10^4$ GeV, $\sigma_1= -141.4$ GeV, $\sin\alpha=0.9$ &  $M_{xd_1}=850$ GeV, $M_{x\ell_i}=500$ GeV, $m_{s_h}=600$ GeV,  $m_{a_h}=2$ TeV, $m_{t_h}=3$ TeV &(0.006, 0.994) & 56.4 fb \\
        \hline
  \end{tabular}
       \caption{Two benchmark scenarios. The cross section is evaluated at the 13 TeV LHC. Note that $y_{ii}^{SD}\,=\,\sqrt{2}\,M_{xd_i}/v_s$,  $y_{ii}^{TD}\,=\, 10^{-5}$
       and we fix $M_{xd_2}=M_{xd_3}=1.5$ TeV.}
         \label{tab:benchmark}
\end{table}
 To analyze the signal in our model, we note that the hardness of the jet from the decay $xd_1 \rightarrow s_h \, j$ will depend on the mass difference between the VLQ $xd_1$ and the scalar 
 $s_h$. This will affect the signal efficiency in the $2\gamma +4j$ channel as well as dictate how well the mass reconstruction for the parent particles can be made. We will 
 discuss these features by considering two benchmark scenarios with different mass gaps between the $xd_1$ and $s_h$, where in one case the jet is hard while for the other case 
 the jet would be comparatively soft.  We choose BP1 ($M_{xd_1}=640$ GeV, $m_{s_h}=600$ GeV) with small mass difference 
between the VLQ and $s_h$ 
 as well as BP2 ($M_{xd_1}=850$ GeV, $m_{s_h}=600$ GeV) with a large mass difference of $250$ GeV between them.
 For the two benchmarks the masses of the VLQs from the other two generations have been kept at 1.5 TeV. 
 By comparing the cross sections for an 850 GeV VLQ  with that of  a 1.5 TeV VLQ  from Fig.~\ref{fig:xd_production} it can be concluded that the two VLQs having mass of 1.5 TeV will 
 not contribute to the final state of the analysis. Other benchmark details including the pair production cross section for $xd_1$ and the branching fractions for the decay of $s_h$ 
 are shown in Table \ref{tab:benchmark}.
 We also check that the current upper limits on the cross section for the diphoton production through a narrow-width scalar resonance at 13 TeV run of LHC given by CMS 
 Collaboration \cite{Khachatryan:2016yec} is satisfied for our choice of the benchmark points. The upper limits on the cross section for the dijet production through 
 a narrow-width resonance at the 13 TeV LHC, given by the CMS collaboration\cite{Sirunyan:2016iap,CMS_dijet_13TeV} are also satisfied.
           
Note that for BP1 the mass difference between $xd_1$ and $s_h$ is 40 GeV. This would mean that the jet coming from the decay of $xd_1$ is quite soft. Although at a hadronic machine
such as the LHC, the jet multiplicity from parton showering would be invariably increased, we intend to focus on relatively hard jets and therefore would like to neglect soft jets in the process.           
So for the analysis of BP1 we consider a final state with smaller jet multiplicity given by $2\gamma \,+ \, \geq \,2j$. We demand that the jets have at least  a minimum 40 GeV transverse momentum. 
The dominant SM background for such a final state is through all subprocesses contributing to $pp\rightarrow 2\gamma \,+ \, \geq \,2j$ (with $p_T(j) \, > \, 40$ GeV).  For BP2 where the mass gap 
between $xd_1$ and $s_h$ is above 200 GeV, one expects the jet from $xd_1$ decay to be quite hard and thus the SM background is given by 
$pp \rightarrow 2\gamma \,+ \, \geq \,4j$.

We have implemented the TeV-scale $U(1)^\prime$ extended model derived from 
$E_6$ GUT in {\tt LanHEP} \cite{lanhep} to generate the model files for {\tt CalcHEP} \cite{Belyaev:2012qa}. 
Using the model files we generated events for the pair production of VLQs (in {\tt LHEF} format \cite{Alwall:2006yp}) at the LHC with $\sqrt{s}=13$ TeV and the subsequent decays of the
$xd_1$ and $s_h$ were included as a decay table for the model (in {\tt SLHA} format\cite{slha}) with the help of {\tt CalcHEP}. We then use these files to decay the unstable particles, and 
pass the generated parton-level events for showering and hadronization in {\tt PYTHIA 8.2} \cite{Sjostrand:2014zea}. 
To enable detector simulation, we then linked the {\tt HepMC2} \cite{Dobbs:2001ck} libraries with {\tt PYTHIA 8.2} to translate {\tt PYTHIA 8} events into {\tt HepMC} format.
For simulating the background, we generated the events at leading-order accuracy using {\tt MadGraph5} \cite{Alwall:2014hca}. {\tt Pythia 6} \cite{Sjostrand:2006za} 
interfaced in {\tt MadGraph5} was used for parton showering and hadronization of the background events, and to get  event files in {\tt STDHEP} format. 

For both signal and background we include the detector effects and have reconstructed the final state objects using {\tt DELPHES 3} \cite{deFavereau:2013fsa}. These are obtained 
in a CMS environment. Further, {\tt FastJet} \cite{Cacciari:2011ma} embedded in {\tt DELPHES} has been used to reconstruct the jets. In the {\tt DELPHES} framework the 
anti-$k_T$ algorithm with a cone size 0.5, $p_T^{j}>20$ GeV and $|\eta_{parton}|<2.5$ is used to reconstruct the jets. The phenomenological event-analysis is done with the 
{\tt MadAnalysis5} package using the event format {\tt ROOT}. 
          
In case of BP1 for which the mass difference between the lightest vector-like quark ($xd_1$) and the scalar ($s_h$) is small, we have generated $pp\rightarrow 2\gamma+2j$ events as background 
at 13 TeV LHC. At the level of generation of events certain basic cuts have been imposed on the final state particles. All jets and photons satisfy $|\eta| < 2.5$ and each final state 
particle is separated from all other final state particles with an angular separation ($\Delta R$) value greater than 0.4. The transverse momenta of photons and jets satisfy 
\begin{align}
p_T(j) \, > \, 20 \,\, \text{GeV}  && \text{and} &&  p_T(\gamma) \,> \, 100 \,\, \text{GeV}.  \label{eq:presel_bp1}
\end{align}

The final state photons for the signal come from the decay of the $s_h$ which has 600 GeV mass and hence the probability for the photons for the signal to have higher $p_T$ values is more compared to the 
background. Hence a 100 GeV $p_T$ cut for photon has been used for the generation of background because the phase space with lower photon $p_T$  will be largely populated by 
background compared to the signal. For 13 TeV LHC, with the above cuts taken into account and at leading-order (LO) accuracy the value of the cross section for the 
parton-level background for BP1 is around 234 fb.

 For BP2 where the mass difference between $xd_1$ and $s_h$ is 250 GeV, $ pp \rightarrow 2\gamma+4j$  events have been generated as the background. The basic cuts on the 
 pseudo-rapidity ($\eta$) and on the angular 
 separation ($\Delta R$) of the final state particles have been taken to be same as that of the benchmark BP1. The cut on the photon $p_T$ is taken to be the 
 same 100 GeV as in case of BP1.  We have imposed different $p_T$ cuts on the four jets and those are given by 
 \begin{align}
p_T(j_1) \, > \, 80 \,\, \text{GeV},  && p_T(j_2) \, > \, 80 \,\, \text{GeV},  && p_T(j_3) \, > \, 40 \,\, \text{GeV} &&   \text{and}  &&  p_T(j_4) \, > \,40 \,\, \text{GeV}.  \label{eq:presel_bp2}
\end{align}
With the above cuts the cross section at the parton-level for the background, for BP2 {\it i.e.} for the process $pp \rightarrow 2\gamma+4j$, comes out to be around 12.15 fb. Similarly the signal events $pp \rightarrow xd_1 xd_1$ have been generated using the event generator {\tt CalcHEP}. The pair production cross sections are shown 
in Table~\ref{tab:benchmark}.
For the reconstructed events we choose the following selection criteria on the photons, jets and leptons
\begin{itemize}
 \item A jet is considered in an event if $p_T(j)>40$ GeV and $|\eta(j)|<2.5$.
  \item An electron or a muon is considered in the lepton set if $p_T(\ell)>10$ GeV and $|\eta(\ell)|<2.5$.
   \item A photon is considered in an event if $p_T(\gamma)>40$ GeV and $|\eta(\gamma)|<2.5$.
   \item All final state candidates are separated from each other with a minimum angular separation satisfying $\Delta R > 0.4$.
\end{itemize}

 \begin{table}[h!]
\centering
\setlength\tabcolsep{4pt}
\begin{minipage}[t]{0.40\textwidth}
\centering
  \begin{tabular}{|l||c|r|}
  \hline
   & \multicolumn{2}{c|}{No. of Events} \\
   \hline
  Cuts    & Signal & Background\\
      \hline
Preselection& 267  &10626 \\
        \hline
    $p_T(\gamma_2)\geq100$ GeV& 248   &9440 \\
        \hline
    $p_T(j_2)\geq100$ GeV& 210  & 2921\\
        \hline
    $p_T(\gamma_1)\geq200$ GeV&205   &1735 \\
        \hline
    $p_T(j_1)\geq150$ GeV& 201  & 1534\\
        \hline
    $M_{eff} \geq  800$ GeV& 201  & 1394\\
        \hline
  \end{tabular}
       \caption{The selected events after each step of selection criteria for BP1 with an integrated luminosity of 100 fb$^{-1}$.}
         \label{tab:table1}
  \end{minipage}%
  \hfill
  \begin{minipage}[t]{0.40\textwidth}
\centering
  \begin{tabular}{|l||c|r|}
   \hline
   & \multicolumn{2}{c|}{No. of Events} \\
  \hline
   Cuts   & Signal & Background\\
      \hline
    Preselection& 40  &685 \\
        \hline
    $p_T(\gamma_2)\geq100$ GeV& 37   &634 \\
        \hline
    $p_T(j_2)\geq100$ GeV& 36  & 554\\
        \hline
    $p_T(\gamma_1)\geq200$ GeV&35   &326 \\
        \hline
    $M_{eff} \geq  1000$ GeV& 35  & 293\\
        \hline
\end{tabular}
  \caption{The selected events after each step of selection criteria for BP2 with an integrated luminosity of 100 fb$^{-1}$.}
    \label{tab:table2}
    \end{minipage}
\end{table}

The preselection criteria for  BP1 is to consider events having 2 photons and a minimum of two jets. 
For BP2 the preselection criteria is 
 to consider events with two photons and a minimum of four jets in the final state.
 We vetoed all the events having at least one isolated lepton of $p_T$ value greater than 
 10 GeV. For the two leading (in $p_T$) jets and the two leading photons in the signal it 
 is expected that they come from the decay of the VLQ. As the mass of the VLQ for the two 
 cases is above 600 GeV, the two leading jets will have a large amount of $p_T$ compared 
 to the leading jets in the background. 
 So for both the benchmarks  we have considered the events with the leading  two jets and 
 two photons with $p_T$ value greater than 100 GeV. 

To further increase the signal-to-background ratio  we apply the following selection cuts to the analysis for BP1   
\begin{align}
   p_T(\gamma_1)\geq 200 \, \text{GeV},  &&  p_T(j_1)\geq 150 \, \text{GeV},  && M_{eff}\geq 800 \, \text{GeV}. 
   \label{eq:selbp1}
\end{align} 
The cut-flow table for BP1 signal and background is shown in Table \ref{tab:table1}. Note that to generate 
the background events with large statistics we have used the preselection cuts given in Eq.~(\ref{eq:presel_bp1}).  
   
Similarly for BP2 we have besides the criterion in Eq.~(\ref{eq:presel_bp1}), 
the additional preselection requirements for jets as given by Eq.~(\ref{eq:presel_bp2})
to generate the SM background with good statistics. We further impose the stronger 
selection cuts on the events  
\begin{align}
 p_T(\gamma_1)\geq 200 \, \text{GeV}, && M_{eff}\geq 1000 \,  \text{GeV}, \label{eq:selbp2}
\end{align}
which help in improving the signal-to-background ratio for the signal events in the 
$2\gamma + 4j$ final state.

With these cuts and a  100 fb${^{-1}}$ integrated luminosity  we get a statistical significance 
of 5$\sigma$ for BP1 as can be seen from the Table \ref{tab:table1}. For BP2 we get a 
significance of 1.9$\sigma$  as can be seen from Table \ref{tab:table2}. The 
significance is calculated using the formula $\sigma = \sqrt{\frac{S}{S+B}}$.
   
 \begin{figure}[h!]
 \centering
 \includegraphics[width=3.2in,height=2.4in]{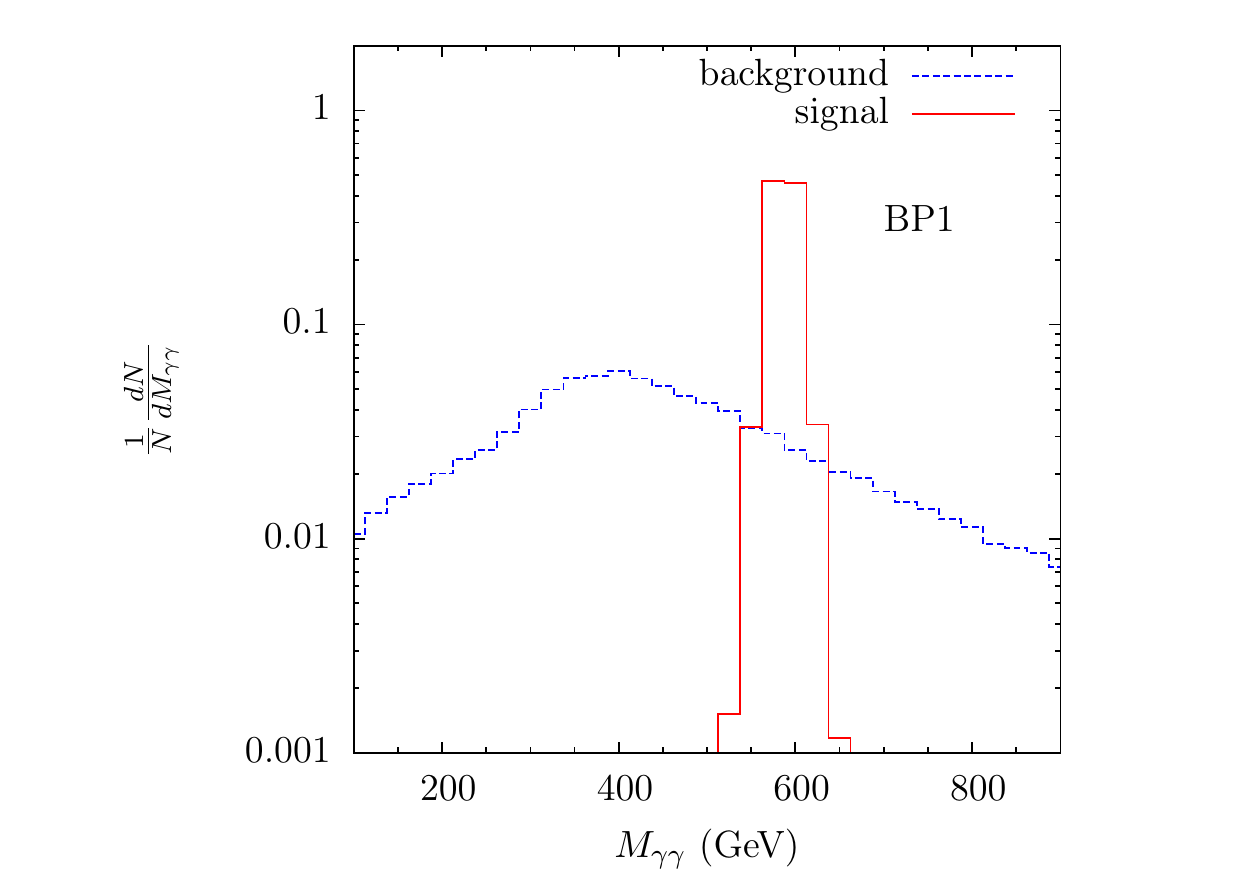}
 \includegraphics[width=3.2in,height=2.4in]{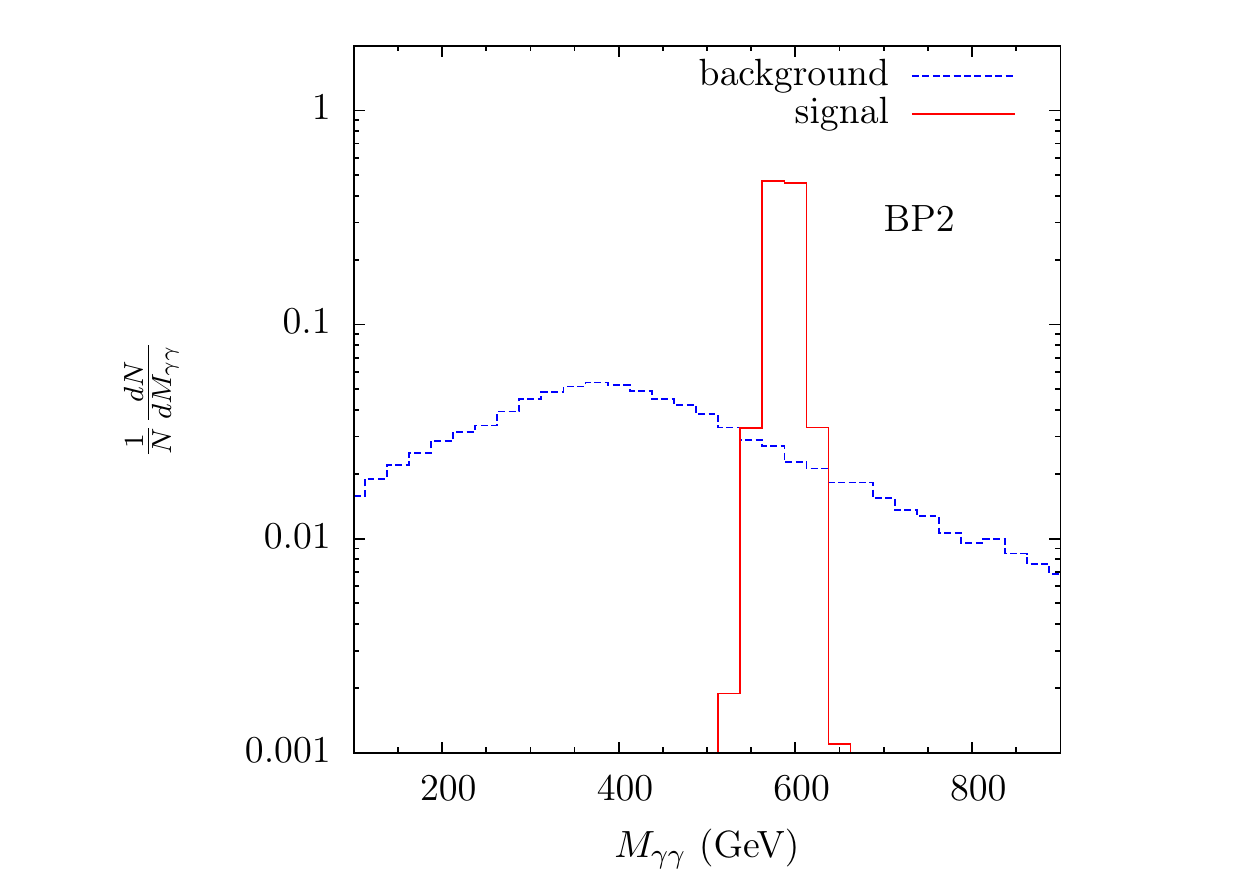}
\caption{The normalized invariant mass distributions for the two leading photons of benchmark 
 scenarios BP1 and  BP2.}
\label{fig:Ma1a2}
\end{figure}
As we trigger upon two hard photons in the final state which come from the decay of the 
scalar $s_h$,  the mass of the scalar $s_h$ can be reconstructed by looking at the invariant
 mass distribution of the two leading photons for both BP1 and BP2. To highlight this 
 we plot the normalized invariant mass distribution of the two leading photons 
 for both signal and background in Fig. \ref{fig:Ma1a2}. 
 As expected the signal from the pair production of the VLQ is confined to a bin around the 
 mass of the scalar $s_h$ with a clear peak for the signal at $m_{s_h}=600$ GeV. There would 
 in principle also be an invariant mass peak for a jet pair around the $s_h$ mass which 
 however is more challenging to observe due to the large spread in their invariant mass 
 distribution.   
  
\begin{figure}[t!]
\centering
\includegraphics[width=3.2in,height=2.4in]{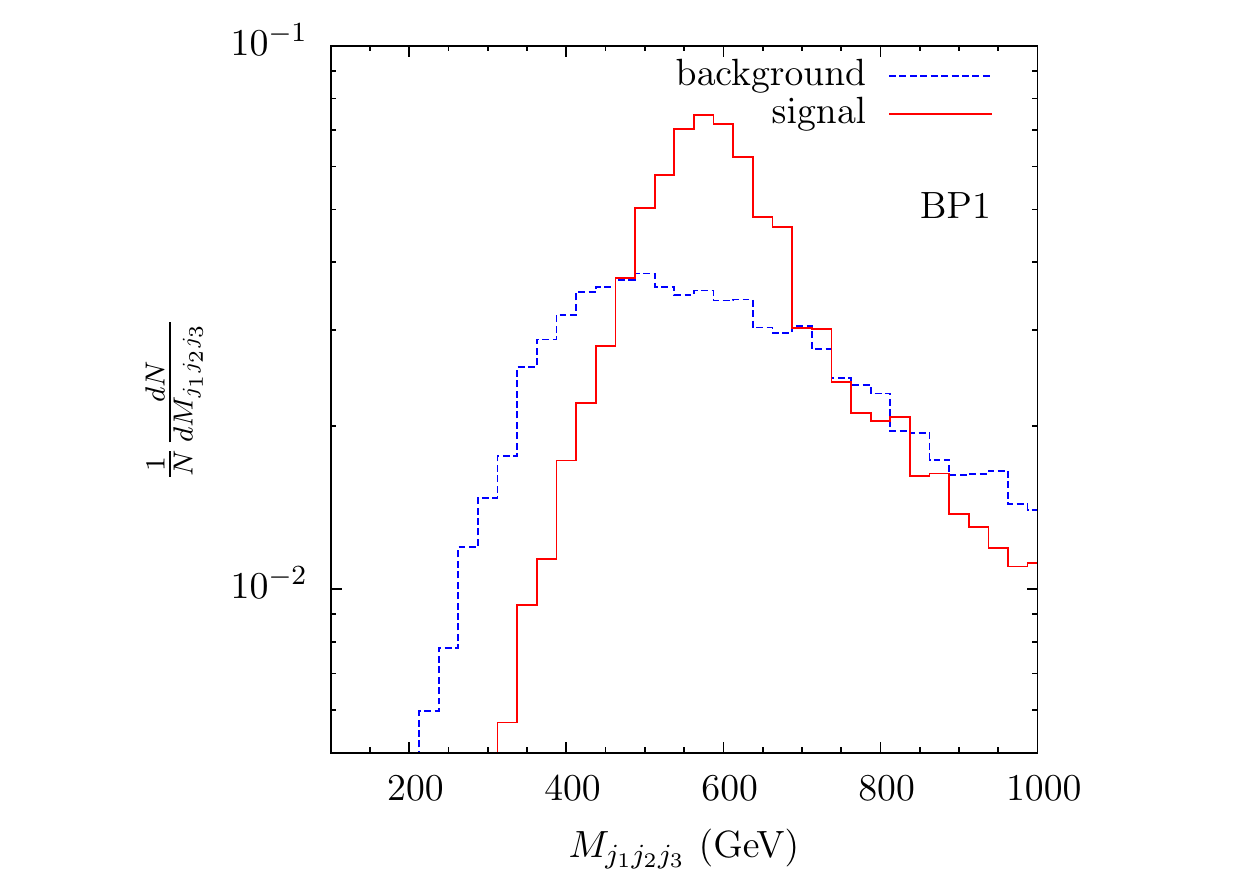}
\includegraphics[width=3.2in,height=2.4in]{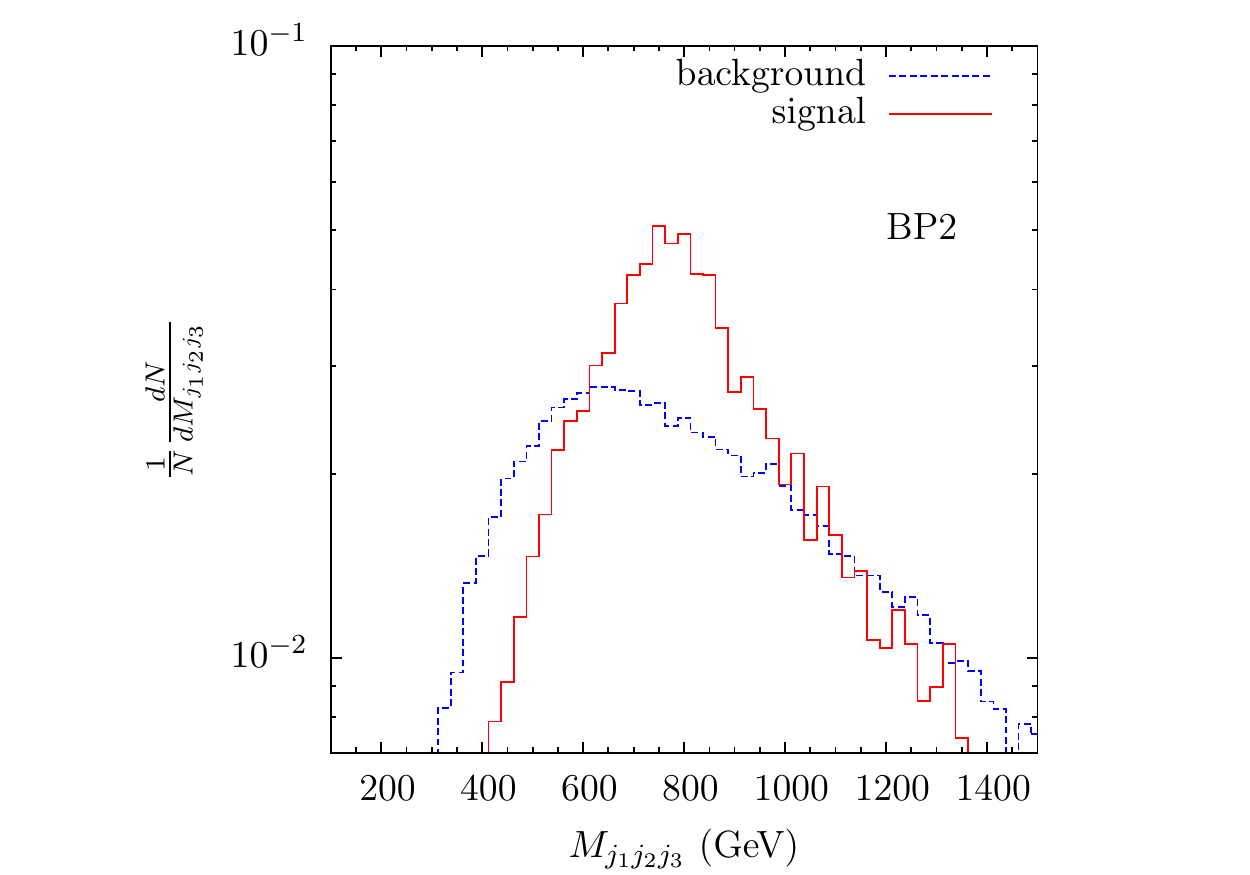}
\caption{The invariant mass distribution for the leading three jets for both benchmark  scenarios BP1 and BP2. }
\label{fig:m3j}
\end{figure}

Similarly, to reconstruct the mass of the VLQ one can use the fully hadronic channel 
giving three jets or the semi-hadronic channel giving two photons and a jet.  
With the knowledge of the reconstructed mass for $s_h$ through the $2\gamma$ 
invariant mass peak, the mass for the vector-like quark can be reconstructed for both 
BP1 and BP2. To compare the reconstruction of the VLQ in the two channels, we first 
plot the $3j$ invariant mass distribution comprised of the leading jets in the events for both
the signal and background in Fig. \ref{fig:m3j}. Although a distinct excess in the distribution
exists around the mass of VLQ for both BP1 and BP2, the spread is quite wide and 
hence unclear as a mass resonance. 

Although, the other channel with two photons and a hard jet should be a much more cleaner 
and precise mode to reconstruct the parent VLQ mass, it does suffer from the ambiguity 
of pairing the right jet with the pair of photons. In addition, for BP1 the mass splitting 
between the $xd_1$ and $s_h$ is quite small and therefore the choice of the right jet is
affected by other soft jets that may originate from showering and fragmentation effects. 
To account for this ambiguity, we use the primary information on the kinematic 
characteristics of events for both BP1 and BP2 that is available to us to 
determine how we should combine the jets with the two photons. Owing to the small mass 
gap in BP1, we can safely assume that the two leading jets for the signal in BP1 would come 
from the decay of $s_h$ and therefore can be safely discounted in the combination. Of the
remaining soft jets, all wrong combinations would only contribute in smearing the distribution
for $M_{\gamma_1\gamma_2 j}$. We therefore propose to neutralize the smearing effects 
by averaging over all such soft jets (with $p_T>40$ GeV) in the invariant mass 
\begin{figure}[h!]
 \centering
  \includegraphics[width=3in,height=2.25in]{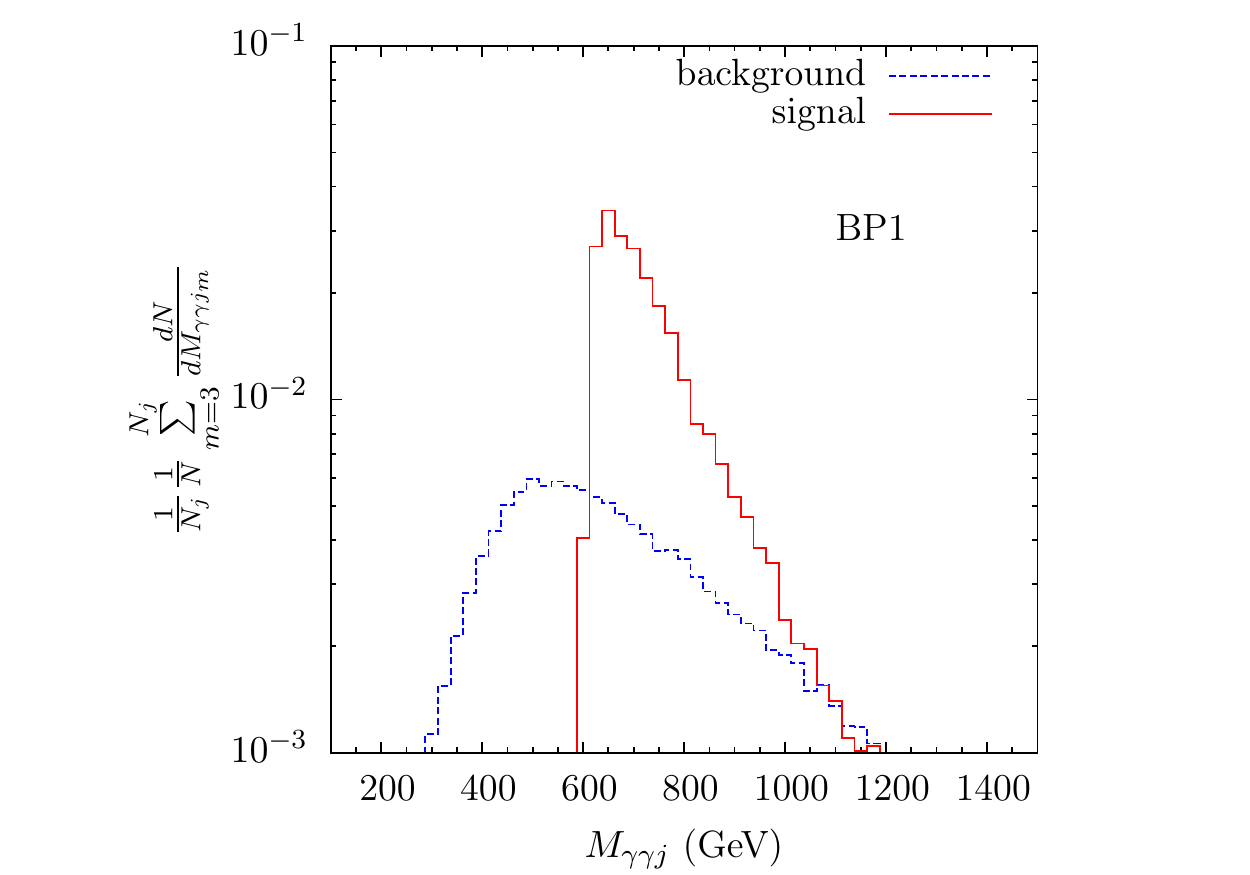}
  \includegraphics[width=3in,height=2.25in]{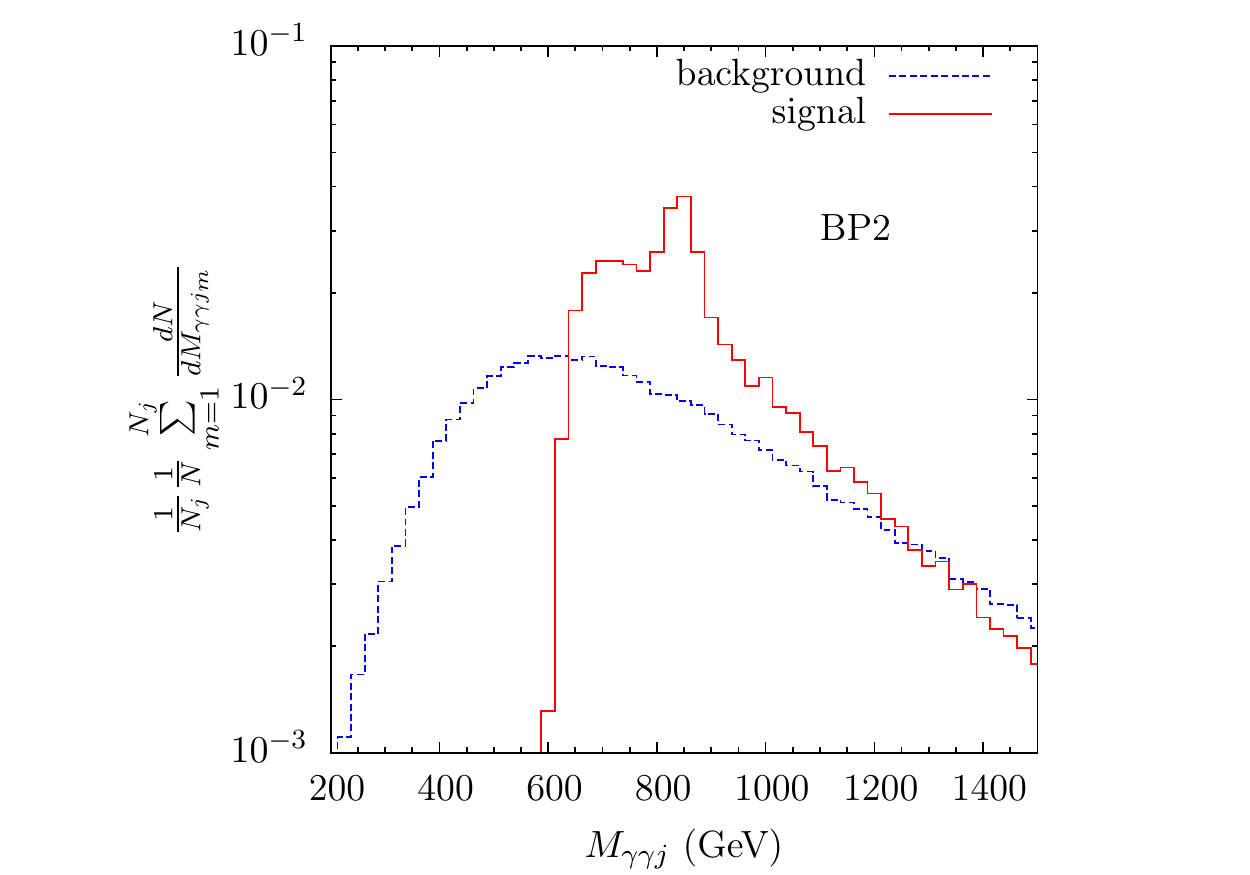}
  \caption{The average of normalized invariant mass distributions of the leading two 
  photons with a jet. The average for  BP1 starts from the third leading jet and the average  
  for BP2 starts from the first leading jet. }
  \label{fig:average_m2aj}
 \end{figure}
reconstruction and neglecting the first two leading jets for BP1. For BP2, the 
jets coming from the decay of VLQ to $s_h j$ is equally hard as the ones that 
come from the decay of $s_h$ themselves. Therefore, for BP2, the averaging is done 
including all jets with $p_T >40$ GeV.  We plot their normalized distribution after 
averaging for both signal and background in Fig. \ref{fig:average_m2aj}. As 
the diphoton coming from the $s_h$ decay marks a kinematic edge in the distribution, this
can be clearly seen to happen at the mass value of $s_h$ at 600 GeV for both BP1 and
BP2 which is absent for the invariant mass distribution in the $3j$ hadronic channel. In
addition, a much cleaner and distinct peak can be observed for the VLQ mass for BP2. 
In case of BP1, as the VLQ mass at 640 GeV is quite close to the scalar $s_h$ mass of 600
GeV, resolving the VLQ (although visible) mass peak  from the sharp kinematic edge is 
difficult. However, for a larger mass gap the peak should be distinctly identifiable as in BP2. 
 
\begin{figure}[h!]
 \centering
 \includegraphics[width=3.5in,height=2.5in]{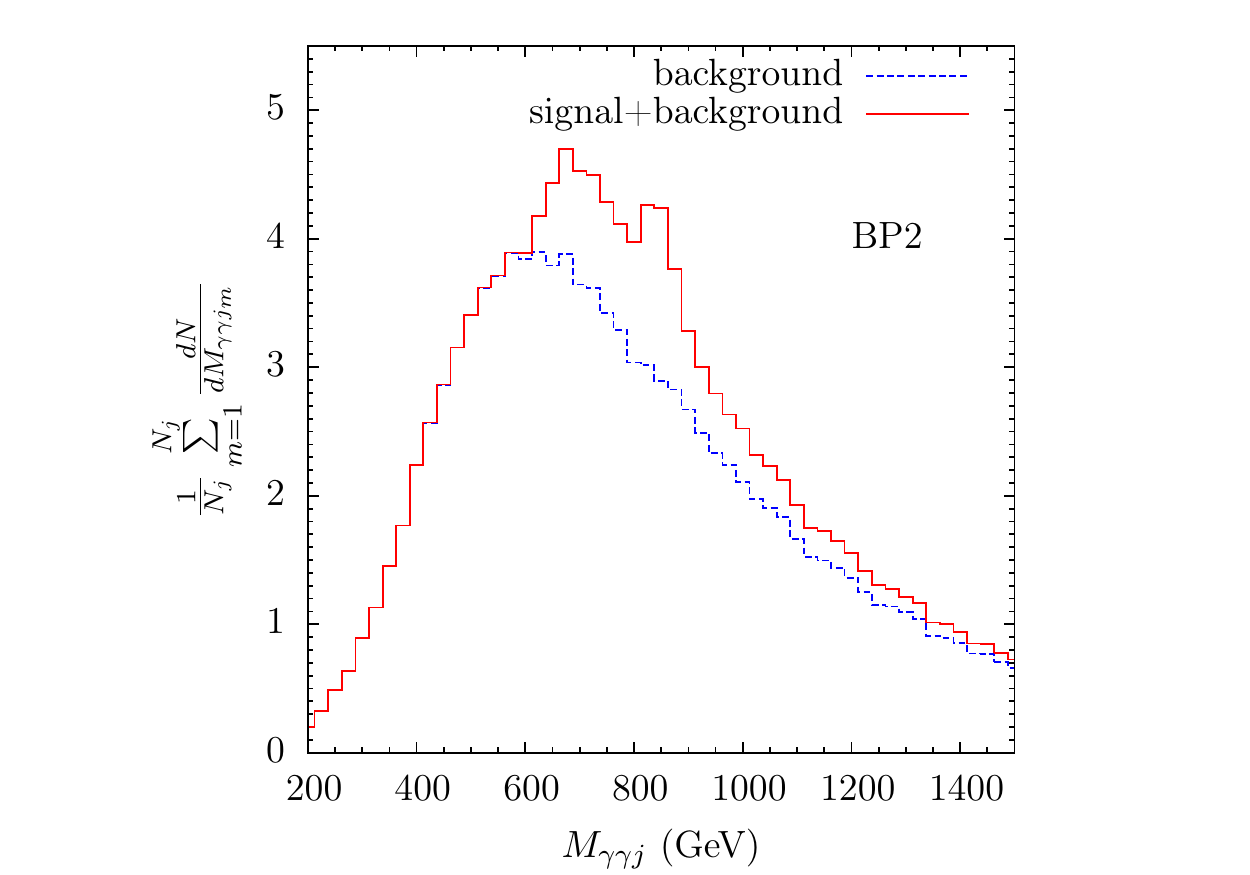}
 \caption{The average of  invariant mass distributions of the leading two  photons with a jet 
 for BP2. }
  \label{fig:average_Ma1a2j}
 \end{figure}

Finally to impress upon the fact that the VLQ mass can be clearly reconstructed through 
the modified invariant mass variable proposed above, we show the distribution without any 
normalization in Fig. \ref{fig:average_Ma1a2j} overlaying the signal for BP2 over the SM 
background. It clearly shows the VLQ mass peak over the background. Thus, we 
find that both the $s_h$ and $xd_1$ can be reconstructed clearly to determine their masses
for the channel under study. As BP2 pertains to a VLQ mass of 850 GeV, we conclude that
a TeV mass VLQ with such non-standard decay modes, possible for BSM scenarios 
which have very little mixing with the SM sector, can be observed and its mass parameters 
determined  at the LHC with a few 100 fb$^{-1}$ of integrated luminosity.


\section{Conclusion}\label{sec:summary}
In this work we have considered an $E_6$ motivated extension of the SM where the 
larger symmetry groups are broken at a very high scale and a residual $U(1)$ gauge 
symmetry is the only remaining symmetry beyond the unbroken SM gauge symmetry. This 
additional $U(1)$ then gets broken at the TeV scale through new SM singlet scalars giving 
rise to a TeV scale particle spectrum with three generations of vector-like quarks and leptons 
and several neutral scalars. The vector-like quarks in the model have non-standard decay 
modes and decay into an ordinary light quark and a SM singlet scalar. Further the 
scalar decays either to two photons or two gluons. The current experimental limits 
for VLQ which do not decay directly to the SM particles are very weak and therefore allow 
their mass to be as light as 500 GeV. We analyzed the events from such VLQ production 
at the LHC with $\sqrt{s}=13$ TeV in the $2\gamma+\geq2j$ final states and present a search 
strategy for observing its signals.  We also studied how to reconstruct the masses 
for both the scalar as well as the VLQ through a modified construction of the invariant mass
variable using the $\gamma\gamma j$ sub-system.
We saw that the mass of the scalar can be reconstructed from the invariant mass distribution 
of the two leading photons. With the upcoming high luminosity data at the LHC, the 
new signal for the VLQ, proposed in this work, could provide to be an interesting channel 
to search for new physics beyond the SM.   

\begin{acknowledgments}
KD would like to thank J. Beuria and S. Dwivedi for useful discussions.
This research was supported in part by the Projects 11475238 and 11647601 supported
by National Natural Science Foundation of China, and by Key Research Program of Frontier Science, CAS (TL).
The work of SN was in part supported by the US department of Energy, Grant number DE-SC-0016013. 
The work of KD and SKR was partially supported by funding available from the Department of Atomic 
Energy, Government of India, for the Regional Centre for Accelerator-based Particle Physics (RECAPP), 
Harish-Chandra Research Institute. The authors acknowledge the use of
the cluster computing setup available at RECAPP and at the High Performance Scientific Computing facility at HRI. 

\end{acknowledgments}


\end{document}